\documentclass[aps,pre,preprint,showpacs,superscriptaddress,english]{revtex4}
\usepackage{graphicx,times,color,xspace}
\usepackage{amsmath,amsthm,amssymb}
\usepackage{babel}
\newcommand{\tim}{\ensuremath{\left( t \right)}}
\newcommand{\tis}{\ensuremath{\left( s \right)}}
 	
\begin{document}

\title{Specificity and Completion Time Distributions of Biochemical Processes}

\author{Brian Munsky}
\affiliation{Center for Nonlinear Studies and Computer, Computation and Statistical Sciences Division,
  Los Alamos National Laboratory, Los Alamos, NM 87545 USA}
\affiliation{Contributed Equally}
\author{Ilya Nemenman}
\affiliation{Center for Nonlinear Studies and Computer, Computation and Statistical Sciences Division,
  Los Alamos National Laboratory, Los Alamos, NM 87545 USA}
\author{Golan Bel}
\affiliation{Center for Nonlinear Studies and Computer, Computation and Statistical Sciences Division,
  Los Alamos National Laboratory, Los Alamos, NM 87545 USA}
\affiliation{Contributed Equally}
\email{golanbel@lanl.gov}


\date{\today}

\begin{abstract}
In order to produce specific complex structures from a large set of similar biochemical building blocks, many biochemical systems require high sensitivity to small molecular differences.
The first and most common model used to explain this high specificity is kinetic proofreading,
which has been extended to a variety of systems from detection of DNA mismatch to cell signaling processes.
While the specification properties of the kinetic proofreading model are well known and were studied in various contexts,
very little is known about its temporal behavior. In this work, we study the dynamical properties
of discrete stochastic two branch kinetic proofreading schemes. Using the Laplace transform of the corresponding chemical master equation, we obtain an analytical solution for the completion time distribution. In particular we provide expressions for the specificity and the
mean and the variance of the process completion times. We also show that, for a wide range of parameters a process
distinguishing between two different products can be reduced to a much simpler three point process.
Our results allow for the systematic study of the interplay between specificity and completion times as well as testing the validity of the kinetic proofreading model in biological systems.
\end{abstract}
\maketitle
%\pacs{}
%\keywords{Kinetic Proofreading, Laplace Transform, Chemical Master Equation, Completion Time, Stochastic Chemical Kinetics}

%\maketitle

\section{Introduction}
The strong bias toward the correct assembly of particular molecular constructs, or specificity,
plays a key role in myriad biochemical processes such as DNA assembly, cell signaling, protein folding, and others.
A common model accounting for the almost error free completion of these processes is {\em kinetic proofreading},
which was first suggested to explain the high specificity of protein synthesis \cite{Hopfield:1974}.
Similar motifs are common in various biological processes where multiple error-prone steps generate error-free results.
For example, kinetic proofreading schemes are common in modeling of DNA synthesis, repair and replication
\cite{Yan:1999,Sancar:2004,Goulian}. 
Similar proofreading ideas appear in other contexts such as protein translation
\cite{Hopfield:1974,Blanchard:2004}, molecular transport \cite{NPC},
receptor-initiated signaling
\cite{McKeithan:1995,Rabinowitz:1996,Rosette:2001,Liu:2001,Goldstein:2004,faeder-03},
RNA transcription \cite{Springgate}, and other processes.
 
Various aspects of the kinetic proofreading concept have already been studied.
Hopfield \cite{Hopfield:1974} and Ninio \cite{Ninio-1} demonstrated the possible increases in specificity due to
single step proofreading. Later explorations of similar proofreading models considered the multi-step proofreading
process as a ``black box'', and studied the accuracy achieved by such processes \cite{Freter} as well as the energy
cost and optimal distribution of the proofreading effort along the proofreading chain \cite{Savageau}. 
In \cite{McKeithan:1995} the kinetic proofreading was proposed as a model for the T-cell receptor explaining
the high discrimination between foreign antigen and self antigen with only moderately lower affinity.
In this context the specificity of a multi step process was studied again as well as the time delay between
initial binding and output signal.

In addition to process specificity, the time required to reach this specificity also plays an important role in
biochemical processes. A proofreading strategy must be efficient as well as specific.
In different contexts \cite{AZJPGB,Chou:2005,redner:2001,Bel:2005PRL,Bel:2006PRE} it was shown that such completion or first passage
times provide a wealth of information about the underlying systems. Extending these results to kinetic proofreading
suggests that the characterization of the completion time distribution may help researchers to distinguish
between different kinetic models and even support or oppose the existence of kinetic proofreading in specific systems.
Surprisingly the completion time distributions of kinetic proofreading schemes haven't been calculated before.

In this article, we investigate the temporal behavior of different {\em kinetic proofreading} (KPR) schemes.  
We derive the chemical master equation (CME--\cite{vanKampen}) and its transform into the Laplace domain, which provides analytical expressions for the directional and non-directional completion time distribution.
In particular, the zeroth, first and second derivatives of the CME's Laplace transform provide expressions for the specificity, mean and coefficient
of variation of the completion times.  In turn these expressions provide a starting point to examine the tradeoffs between the stationary
and temporal behaviors of different KPR schemes. Furthermore, we show that over a wide range of kinetic parameters
the complex proofreading process reduces to a three-state process with simple distributions of the transition time
between the three states. We also provide a diagram mapping the parameters space into classes of different behavior
of the completion time distribution. This paper is organized as follows.
In Section II, we introduce the model and provide its chemical master equation as well as the analytical solution of the CME in the Laplace domain. In Section III we show the different behaviors of the completion time distributions and divide the 
parameters space into regimes corresponding to different typical distributions.
We also show the coefficient of variation versus the parameters of the problem and discuss it's meaning.
In Section IV we summarize our results and their relevance to many of the problems previously studied 
in the context of kinetic proofreading.    

%------------------------------------------------------------
\begin{figure}[t]
\includegraphics[width=0.9\linewidth]{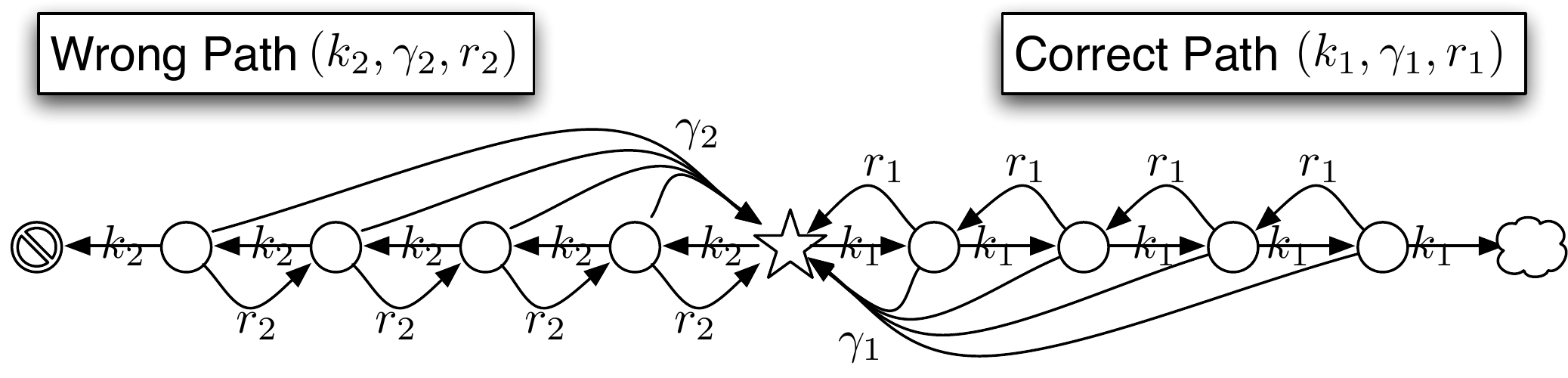}
\caption{Schematic description of the two-branch general kinetic proofreading scheme for error correction.
The process begins at the point denoted with a star. From there it can hop right or left one jump at a time
with rate $k_1+k_2$. On the right half of the chain, the process can continue one step forward with rate $k_1$,
it can also move one step backward with rate $r_1$ or return to the initial point with rate $\gamma_1$.
On the left half of the chain, these rates are replaced with $k_2,r_2$ and $\gamma_2$.
The leftmost and rightmost sites are absorbing sites, once the particle reaches these points, the process is completed.
If the particle finishes at the rightmost site, the process is said to have completed correctly, if it finishes
at the leftmost site, the process has completed incorrectly.}
\label{fig:Model2}
\end{figure}
\section{The Model}
Here we consider the general model of kinetic proofreading (KPR), which can be represented by the Markov chain
in Fig.\ \ref{fig:Model2}.  The initiation state is represented by the star in the center of the chain,
and is denoted by $(i,j)=(0,0)$. Depending upon the system, the state $(i,j)=(0,0)$ may have different meanings;
in protein assembly this state may correspond to an empty A-site of the mRNA-ribosome complex \cite{Hopfield:1974},
or in cell signaling the initiation state may correspond to a receptor with no bound ligand \cite{McKeithan:1995}.
The state just to the right of the star, labeled by $(i,j)=(1,0)$ corresponds to a single step in the ``correct"
direction, i.e. the intended tRNA binds to the A-site or the proper ligand binds to the receptor.
Conversely, a step to the left is in the wrong direction (wrong tRNA or wrong ligand).
In general there may be many wrong directions or additional sub-chains branching from the central initiation point,
but for simplicity we consider only the case where there is only one right and one wrong decision.
The Markov system can transition one step away from the initiation point with rate $k_1$ in the correct direction
or $k_2$ in the incorrect direction. The process may also move one step toward the initiation point with rate $r_1$
or $r_2$, or back to the origin with rate $\gamma_1$ or $\gamma_2$. The two branches of the chain have $L_1$ or $L_2$ nodes
correspondingly, the last of which, $(L_1,0)$ or $(0,L_2)$ is an absorbing point (representing the formation of the
relevant final product).
The chemical master equation (CME) describing the dynamics of the occupation probabilities is:
{\footnotesize \begin{widetext}
\begin{align}
\hspace{-20pt}\frac{dp_{i,j}\tim}{dt}&=\left\{ \begin{array}{ll} 
k_2p_{0,L_2-1}\tim&\text{  for }(i,j)=(0,L_2)\\
-\left( k_2+\gamma_2 +r_2 \right) p_{0,L_2-1}\tim+k_2p_{0,L_2-2}\tim &\text{  for }(i,j)=(0,L_2-1) \\ 
-\left( k_2+\gamma_2 +r_2 \right) p_{0,j}\tim+k_2p_{0,j-1}\tim+r_2p_{0,j+1}\tim& \text{  for $i=0$ and }0<j<L_2-1 \\ 
-(k_1+k_2)p_{0,0}\tim+r_1p_{1,0}\tim+r_2p_{0,1}\tim+\gamma_1 \displaystyle{\sum\limits_{i=1}^{L_1-1}}p_{i,0}\tim+\gamma_2 \displaystyle{\sum\limits_{j=1}^{L_2-1}}p_{0,j}\tim  &\text{  for } (i,j)=(0,0) \\
-\left( k_1+\gamma_1 +r_1 \right) p_{i,0}\tim+k_1p_{i-1,0}\tim+r_1p_{i+1,0}\tim& \text{  for $j=0$ and }0<i<L_1-1 \\ 
-\left( k_1+\gamma_1 +r_1 \right) p_{L_1-1,0}\tim+k_1p_{L_1-2,0}\tim &\text{  for }(i,j)=(L_1-1,0) \\
k_1p_{L_1-1,0}\tim&\text{  for }(i,j)=(L_1,0).
\end{array}\right.  \label{ME}
\end{align}
\end{widetext}}
For any given specific case, this CME may be solved using various methods, such as various projection approaches \cite{Munsky:2005FSP,Burrage:2006,Munsky:2007mtsFSP,Slaven:2006JCP,Munsky:2008IEEE}, or simulated using stochastic simulations \cite{Gillespie:1976,Gillespie:2001,Petzold:2005}.  Similarly, completions times for a given process could be calculated directly from the CME using projection approaches \cite{Munsky:2008IET} or analyzed using transition path and transition interface sampling \cite{Dellago:1998,Faradjian:2004,Moroni:2004,Erp:2005,Allen:2006JCPA}.  However, in this work we take an analytical approach in an effort to attain explicit expressions for the temporal behavior of the process in terms of the kinetic parameters. Later in Section III, those explicit expressions will better enable us to study the dependence of the specificity and completion time distributions on the system's parameters as the number of intermediate steps, and forward/backward/proofreading rates. More specifically, we first simplify the set of differential equation describing the dynamics of the occupation probabilities, by applying the Laplace transform:
%~~~~~~~~~~~~~~~~~~~
\begin{equation}
P_{i,j}\tis\equiv\displaystyle{\int\limits_{0}^{\infty}}p_{i,j}\tis e^{-st}dt,
\end{equation}
%~~~~~~~~~~~~~~~~~~~~~~~~~~~
where we are using lowercase variables to represent quantities in the time domain and uppercase variables to
represent the corresponding quantities in the Laplace domain.
Upon application of the Laplace transform, the probabilities are now described by the following 
algebraic master equation
%~~~~~~~~~~~~~~~~~~~~~~~~~~~~~~~~~~~~~~~~~~~~~
{\footnotesize \begin{widetext}
\begin{align}
\hspace{-15pt}P_{i,j}\tis&=\left\{ \begin{array}{ll} 
\frac{k_2}{s}P_{0,L_2-1}\tis  &\text{  for }(i,j)=(0,L_2)\\ 
\frac{k_2}{s+k_2+\gamma_2 +r_2}P_{0,L_2-2}\tis  &\text{  for }(i,j)=(0,L_2-1)  \\ 
\frac{1}{s+k_2+\gamma_2 +r_2}\left(k_2P_{0,j-1}\tis +r_2P_{0,j+1}\tis\right) &\text{  for $i=0$ and }0<j<L_2-1  \\ 
\frac{1}{s+k_1+k_2}\left(1+r_1P_{1,0}\tis+r_2P_{0,1}\tis+\gamma_1 \displaystyle{\sum\limits_{i=1}^{L_1-1}}P_{i,0}\tis +\gamma_2 \displaystyle{\sum\limits_{j=1}^{L_2-1}}P_{0,j}\tis \right)&\text{  for } (i,j)=(0,0) \\ 
\frac{1}{s+k_1+\gamma_1 +r_1}\left(k_1P_{i-1,0}\tis +r_1P_{i+1,0}\tis\right) &\text{  for $j=0$ and }0<i<L_1-1  \\ 
\frac{k_1}{s+k_1+\gamma_1 +r_1}P_{L_1-2,0}\tis  &\text{  for }(i,j)=(L_1-1,0)  \\ 
\frac{k_1}{s}P_{L_1-1,0}\tis  &\text{  for }(i,j)=(L_1,0) .
 \label{MELaplace}
\end{array}\right.
\end{align}
\end{widetext}}
For the above equation we have already imposed the initial condition $p_{i,j}\left(t=0\right)=\delta_{i,0}\delta_{j,0}$,
where $\delta$ is the Kronecker delta. In other words, $p_{0,0}(0)=1$ and $p_{i,j}(0)=0$ for all $(i,j)\neq(0,0)$.
The general solution of these equations is explicitly written as
%~~~~~~~~~~~~~~~~~~~~~~~~~~~~~~~~~~~~~~~~~~~~~~~~~~~~~~~~~~~~~~
\begin{align}\label{pssol}
P_{i,j}\tis =\left\{\begin{array}{ll} 
A\lambda_1^i+B\lambda_2^i  &\text{  for $j=0$, $i\geq 0$}\\ 
A\beta_2^j+B\beta_2^j+C(\beta_1^j-\beta_2^j) &\text{  for $i=0$, $j> 0$}\\ 
\end{array}\right.
\end{align}
%~~~~~~~~~~~~~~~~~~~~~~~~~~~~~~~~~~~~~~~~~~~~~~~~~~~~~~~~~~~~~~~~~~~~~~
The space independent parameters $\lambda_{1,2}(s)$ and $\beta_{1,2}(s)$ are obtained from the solution of
the quadratic equations
%~~~~~~~~~~~~~~~~~~~~~~~~~~~~~~~~~~~~~~~~~~~~~~~~~~~~~~~~~~~~~~~~~~~~
\begin{align}
\frac{k_1}{s+k_1+\gamma_1 +r_1}+\frac{r_1}{s+k_1+\gamma_1 +r_1}\lambda^{2}-\lambda &=0 \nonumber \\ 
\frac{k_2}{s+k_2+\gamma_2 +r_2}+\frac{r_2}{s+k_2+\gamma_2 +r_2}\beta^{2}-\beta &=0, \label{quad}
\end{align}
%~~~~~~~~~~~~~~~~~~~~~~~~~~~~~~~~~~~~~~~~~~~~~~~~~~~~~~~~~~~~~~~~~~~~~~~~~
which come from the expressions for $P_{i,j}(s)$ at the interior points of the two branches.  
The boundary conditions are satisfied by proper choice of the coefficients $A(s), B(s)$ and $C(s)$.
The boundary condition at $(i,j)=(0,0)$ (see Eq.~\ref{MELaplace}) is expressed as:
\begin{align}
(s+k_1+k_2)(A+B) &=1+r_1(A\lambda_1+B\lambda_2)+r_2((A+B)\beta_2+C(\beta_1-\beta_2)) \nonumber \\ &+\gamma_1 \displaystyle{\sum\limits_{i=1}^{L_1-1}}\left(A\lambda_1^i+B\lambda_2^i\right) +\gamma_2 \displaystyle{\sum\limits_{j=1}^{L_2-1}}\left((A+B)\beta_2^j+C(\beta_1^j-\beta_2^j)\right) \label{BC1}.
\end{align}
The boundary condition at $(i,j)=(L_1-1,0)$ is written as:
\begin{align}
A\lambda_1^{L_1-1}+B\lambda_2^{L_1-1}&=\frac{k_1}{{s+k_1+\gamma_1+r_1}}\left(A\lambda_1^{L_1-2}+B\lambda_2^{L_1-2}\right), \label{BC2}
\end{align}
and the boundary condition at $(0,L_2-1)$ is
\begin{widetext}
\begin{align}
A\beta_2^{L_2-1}+B\beta_2^{L_2-1}&=\frac{k_2}{s+k_2+\gamma_2 +r_2}\left(A\beta_2^{L_2-2}+B\beta_2^{L_2-2}+C(\beta_1^{L_2-2}-\beta_2^{L_2-2})\right)-C(\beta_1^{L_2-1}-\beta_2^{L_2-1}). \label{BC3} 
\end{align}
\end{widetext}
Using the definitions of $\lambda_{1,2}$ (see Eq.~\ref{quad}) we can rewrite Eq.~(\ref{BC2}) as
\begin{align}
B=-A\frac{\lambda_1^{L_1}}{\lambda_2^{L_1}}. \label{BC2b} 
\end{align}
Similarly using the definitions of $\beta_{1,2}$ we rewrite Eq.~(\ref{BC3}) as
\begin{align}
C%&=A\left(1-\frac{\lambda_1^{L_1}}{\lambda_2^{L_1}}\right)\Bigg/\left(1-\frac{\beta_1^{L_2}}{\beta_2^{L_2}}\right)\\
&=A\frac{\beta_2^{L_2}\left(\lambda_2^{L_1}-\lambda_1^{L_1}\right)}
{\lambda_2^{L_1}\left(\beta_2^{L_2}-\beta_1^{L_2}\right)}. \label{BC3b} 
\end{align}
Finally, using Eqs.~(\ref{BC2b},\ref{BC3b}) one can simplify Eq.~(\ref{BC1})
\begin{widetext}
\begin{align}\label{BC1b}
\frac{1}{A}=&  \left(1-\frac{\lambda _1^{L_1}}{ \lambda _2^{L_1}}\right) \left(\gamma_2+k_1+k_2+s+\gamma_1-\gamma_2\frac{\frac{1-\beta_2^{L_2} }{1-\beta_2}\beta_1^{L_2}+\frac{1-\beta_1^{L_2} }{1-\beta_1}\beta _2^{L_2}}{\beta_2^{L_2}-\beta _1^{L_2}}-r_2\frac{ \beta _2 \beta _1^{L_2}+\beta _1 \beta_2^{L_2}}{\beta _2^{L_2}-\beta _1^{L_2}} \right) \\ \nonumber
&-r_1 \lambda_1 \left(1-\frac{\lambda_1^{L_1-1}}
{\lambda _2^{L_1-1}}\right)-\gamma_1  \left(\frac{1-\lambda _1^{L_1}}{1-\lambda _1}-\frac{\lambda _1^{L_1}}{ \lambda_2^{L_1}}\frac{1-\lambda_2^{L_1}}{1-\lambda _2}\right). 
\end{align}
\end{widetext}
Note that in deriving Eqs.(\ref{BC2b},\ref{BC3b},\ref{BC1b}) we assumed that the parameters
$k_1,k_2,r_1,r_2,\gamma_1,\gamma_2$ are different than zero.

In order to study the temporal behavior of the kinetic proofreading model, we compute (i) the probability that the system will reach
the correct terminus point and (ii) the distribution of time
until the system reaches one of the two possible terminus points. 
Both of these quantities are found by examining the probability density function (PDF) for the first passage time
to the absorbing sites $(L_1,0)$ or $(0,L_2)$ which are given by:
%~~~~~~~~~~~~~~~~~~~~~~~~~~~~~~~~~~~~~
\begin{align}
f_{1}\tim&=k_1p_{L_1-1,0}\tim \nonumber\\
f_{2}\tim&=k_2p_{0,L_2-1}\tim.  \label{FPTPDF}
\end{align}
%~~~~~~~~~~~~~~~~~~~~~~~~~~~~~~~~~~~~~~~~~~~~~~
According to Eqs.~(\ref{FPTPDF}) and (\ref{pssol}) the Laplace transform of the first passage time PDF is given by
%~~~~~~~~~~~~~~~~~~~~~~~~~~~~~~~~~~~~~~~~~~~~~~~~
\begin{align}
F_{1}\tis& = k_1\left(A\lambda_{1}^{L_1-1}+B\lambda_{2}^{L_1-1}\right),\nonumber\\
F_{2}\tis& = k_2\left(C\beta_{1}^{L_2-1}+(A+B-C)\beta_{2}^{L_2-1}\right).  \label{LFPTPDF}
\end{align}
%~~~~~~~~~~~~~~~~~~~~~~~~~~~~~~~~~~~~~~~~~~~~~~~~~~~~~~
%--------------------------------------------------------------------------------

%--------------------------------------------------------------------------
This expression now contains a wealth of information about the moments of the escape time distributions.
For example, the probability of reaching the correct absorbing site, $(i,j)=(L_1,0)$, is found by evaluating $F_1(s)$
at $s=0$. Furthermore, the $m^{th}$ moment of the arbitrary completion time is
%~~~~~~~~~~~~~~~~~~~~~~~~~~~~~~~~~~~~~~~~~~~~~~~~~~~~~
\begin{align}
T_{T}^{(m)}&=\displaystyle{\int\limits_{0}^{\infty}t^m(f_{1}\tim+f_{2}\tim) dt} \nonumber \\ 
&=\left(-1\right)^m\left. \left(\frac{dF_{1}\tis}{ds}+\frac{dF_{2}\tis}{ds}\right)\right|_{s=0}, \label{align}
\end{align}
%~~~~~~~~~~~~~~~~~~~~~~~~~~~~~~~~~~~~~~~~~~~~~~~~~~~~~~~~~~~~
and the $m^{th}$ normalized moment of the escape time to the correct site $(i,j)=(L_1,0)$ is:
%~~~~~~~~~~~~~~~~~~~~~~~~~~~~~~~~~~~~~~~~~~~~~~~~~~~~~~~~~~~~~~
\begin{equation}
T_1^{(m)}=\frac{\left(-1\right)^m}{F_1(0)}\left. \left(\frac{dF_{1}\tis}{ds}\right)\right|_{s=0}, \label{moments2}
\end{equation}
%~~~~~~~~~~~~~~~~~~~~~~~~~~~~~~~~~~~~~~~~~~~~~~~~~~~~~~~~~~~~~~~~~~
%------------------------------------------------------------------
%------------------------------------------------------------------------
\section{Results and Discussion}
The non-normalized Laplace transforms of the two branches, $F_1(s)$ and $F_2(s)$ provide a complete description
of the completion process and in particular, we analyze  two important quantities: (1) the probability that the
process completes via one branch or the other and (2) the distribution of time needed for this completion.
In the latter case, we concentrate our attention on the mean and variance of the completion times.
For the general two-branch process, it is relatively simple to generate symbolic expressions for the
completion probabilities and the moments of the completion times.
Where these expressions are simple enough to be informative, we will provide their explicit forms
for which we will use the following notation
\begin{equation}\label{defs}
l_{1,2}=\lambda_{1,2}\vert_{s=0};\ \ \ \ b_{1,2}=\beta_{1,2}\vert_{s=0};\ \ \ \ \text{and} \ \ \ \ A_{0}=A\vert_{s=0}.
\end{equation}
Where the expressions are not sufficiently compact, particularly for the higher moments of the completion
time distributions, we will use numerical examples to illustrate their dependence on parameters.
For these numerical examples, we fix the length of each branch to involve $L_1=L_2=16$ steps.
To explore the effect of different time scales in each branch, we will consider the case when the forward rates
of both branches are equal ($k_1=k_2$) and the case where the forward rate of the ``correct" branch is six times
that of the ``wrong" branch ($k_1=6k_2$).

\subsection{``Correct"  and ``Wrong" Completion Probabilities}

In a kinetic proofreading process, the biochemical process must somehow give preference to completing in the
correct way, i.e.\ adding the correct amino acid to the growing protein chain or initiating intracellular signaling
when the correct ligand is bound to the receptor, but not when the incorrect ligand is bound.
In our simplified model, this preference corresponds to reaching one absorbing site rather than the other.
Here we analyze how changes in the relevant parameters affect this preference.
Following the derivations in the previous section we can write the ``correct" or ``wrong" completion probabilities
($P_{C}$ and $P_{W}$, respectively) as
\begin{align}\label{0-mom}
& P_{C}=F_{1}\left(0\right)=k_1l_1^{L_1-1}\left(1-l_1/l_2\right)A_0, \nonumber \\
& P_{W}=F_{2}\left(0\right)=k_2\left(\frac{1-\left(l_1/l_2\right)^{L_1}}{1-\left(b_1/b_2\right)^{L_2}}b_1^{L_2-1}+\left(1-\left(\frac{l_1}{l_2}\right)^{L_1}-\frac{1-\left(l_1/l_2\right)^{L_1}}{1-\left(b_1/b_2\right)^{L_2}}\right)b_2^{L_2-1}\right)A_0.
\end{align}

For example, one can use these expressions to derive expressions for the directional completion probabilities for the
directed kinetic proofreading (dKPR) scheme ($\gamma_{1,2}>0$ and $r_{1,2}=0$) which are
\begin{align}\label{dkpr-prob}
%\leftexp{dKPR}{P}_{C}&
{P}_{C-{\rm dKPR}}
&=\frac{\left(k_1/k_2\right)\left(1+\psi_2 \right)^{L_2-1}}{\left(1+\psi_1 \right)^{L_1-1}+\left(k_1/k_2\right)\left(1+\psi_2 \right)^{L_2-1}} \nonumber \\
%&=\frac{\left(k_1/k_2\right)\left(1+\psi_1 \right)\left(1+\psi_2 \right)^{L_2}}{\left(1+\psi_2 \right)\left(1+\psi_1 \right)^{L_1}+\left(k_1/k_2\right)\left(1+\psi_1 \right)\left(1+\psi_2 \right)^{L_2}} \nonumber \\
%\leftexp{dKPR}{P}_{W}&
{P}_{W-{\rm dKPR}}
%&=\frac{\left(1+\psi_2 \right)\left(1+\psi_1 \right)^{L_1}}{\left(1+\psi_2 \right)\left(1+\psi_1 \right)^{L_1}+\left(k_1/k_2\right)\left(1+\psi_1 \right)\left(1+\psi_2 \right)^{L_2}}.
&=\frac{\left(1+\psi_1 \right)^{L_1-1}}{\left(1+\psi_1 \right)^{L_1-1}+\left(k_1/k_2\right)\left(1+\psi_2 \right)^{L_2-1}}.
\end{align}
where we have used the notation $\psi_{1,2}=\gamma_{1,2}/k_{1,2}$.

\begin{figure*}[t]
\includegraphics[width=0.7\linewidth]{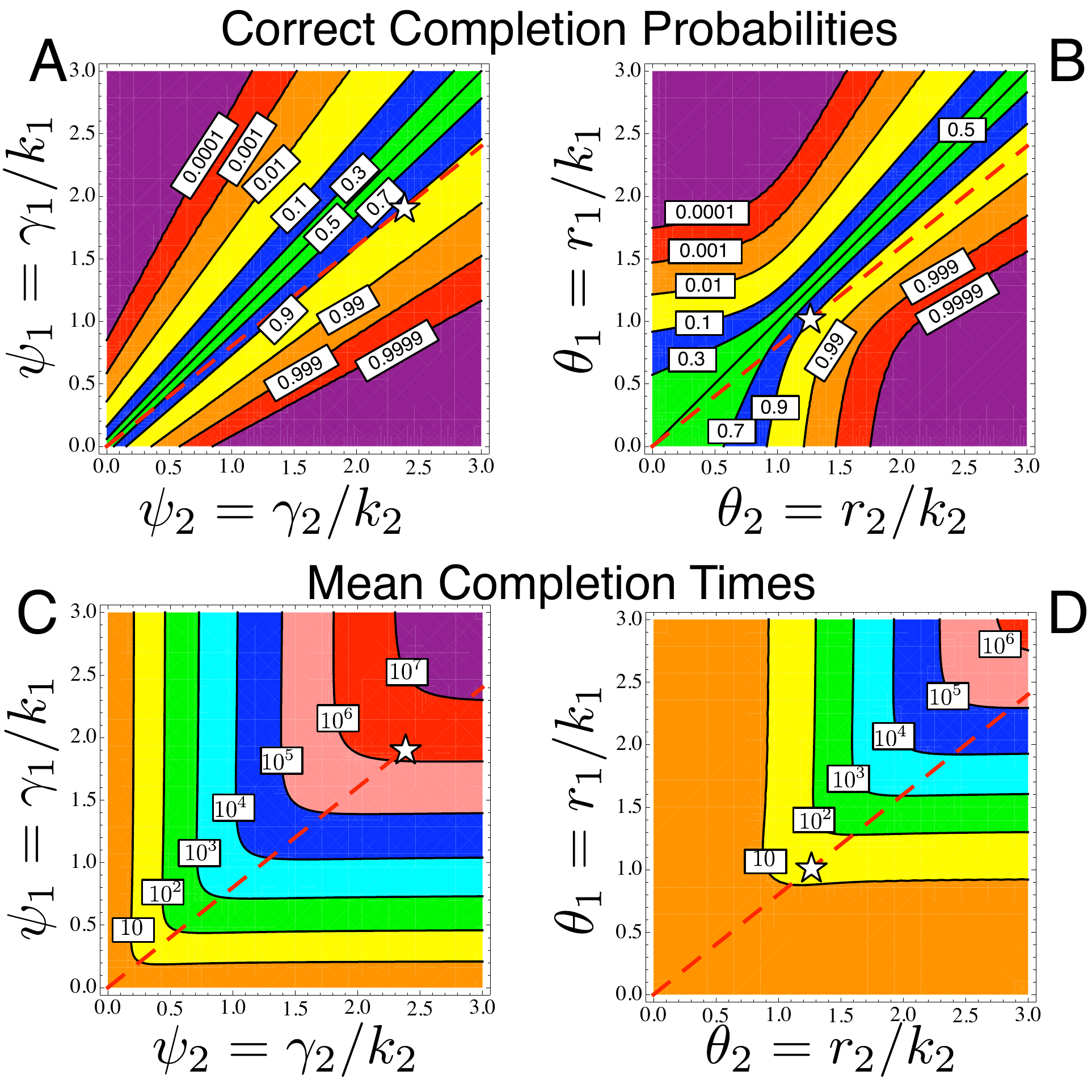}
\caption{
{\em Proofreading with Equal Forward Rates, $k_1=k_2=1$}.
Contour plots of the probability of correct completion (A,B) and the corresponding mean decision time (C,D)
for two different decision processes. (A,C) For the dKPR process with varying kinetic proofreading rates
$\psi_1= \gamma_1/k_1$ and $\psi_2= \gamma_2/k_2$ and zero backward rates, $r_{1,2}=0$.
(B,D) For the AM process with varying backward rates $\theta_1= r_1/k_1$ and $\theta_2= r_2/k_2$ and zero
proofreading rates, $\gamma_{1,2}=0$. For both plots, the lengths of the branches are $L_1=L_2 = 16$, and the contour
lines denote the probabilities of correct completion (upper panels) or mean completion time in units of $1/k_2$
(lower panels). The red dashed line corresponds to a twenty percent difference in the proofreading or backward
{\em ratios}, $\psi_1=0.8\psi_2$ or $\theta_1=0.8\theta_2$, respectively.}
\label{fig:dKPR_Correct}
\end{figure*}

Fig.\ \ref{fig:dKPR_Correct}A shows the probability of completing in the first direction as a function of the kinetic
proofreading rates $\psi_{1,2}$ in the case of equal forward rates ($k_1=k_2=1$).
From the figure, it is apparent that a large amount of specificity is achievable for the properly chosen combination
of $\psi_1$ and $\psi_2$. For example, the system will complete in the correct direction more than 99.99\% percent
of the time for any ($\psi_1,\psi_2$) combination in the lower right corner.
Similarly, one can compute the directional probabilities in the case of the absorption mode (AM) process
(see Fig.\ \ref{fig:dKPR_Correct}B), where $\gamma_{1,2}=0$ but the backward rates $r_{1,2}$ are allowed to vary.
In this case the contour lines for the completion probabilities are less trivial than for the dKPR case.
In particular, the contour lines exhibit a bottle neck near the values of $\theta_{1,2}\equiv r_{1,2}/k_{1,2} = 1$ where the specificity
can change dramatically despite relatively small changes in the parameter values. 

The objective of kinetic proofreading is to provide large amplification in directional specificity despite small
changes in the parameters $\psi$ or $\theta$. To compare how well the dKPR and AM processes achieve this objective
we have drawn red dashed lines in each plot corresponding to $\psi_1 = 0.8\psi_2$ or $\theta_1 = 0.8 \theta_2$,
i.e.,\ there is a twenty percent difference in the relative proofreading or backward ratios, respectively,
between the two branches. Since $k_1=k_2$, this is equivalent to exploring to a 20 percent different in the actual rates
$\gamma$ and $r$. As the backward and proofreading rates increase, the specificity also increases for both process,
as can be seen by how the dashed lines cross the contour levels. The first observation to note is that both the dKPR
and the AM process can attain 90\% specificity with twenty percent difference in rates
(see stars in Figs.\ \ref{fig:dKPR_Correct}A-B) and values of the parameters which are within the range of the plots.

\begin{figure*}[t]
\includegraphics[width=0.7\linewidth]{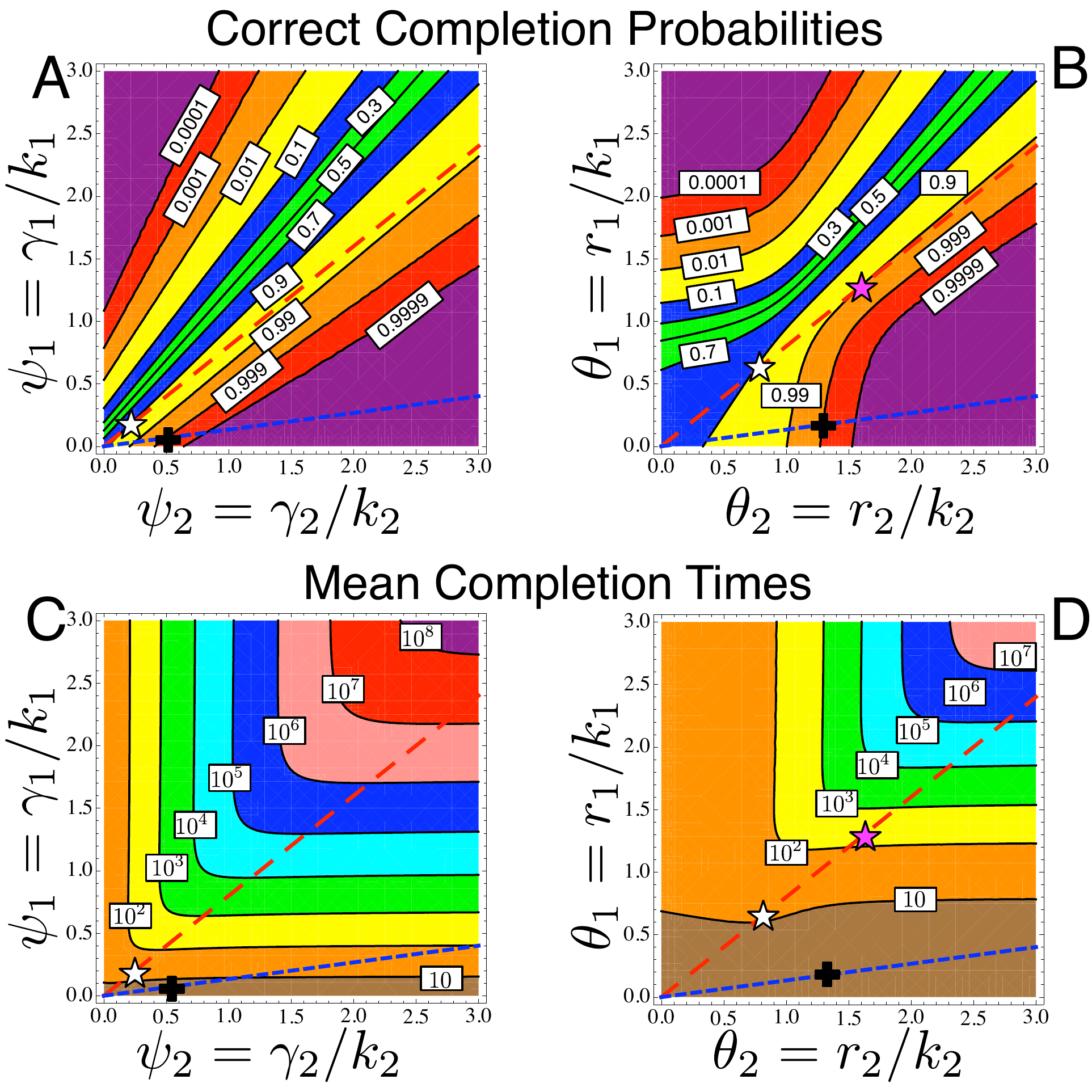}
\caption{
{\em Proofreading with Different Forward Rates, $k_1=6$ and $k_2=1$}.
Contour plots of the probability of correct completion (A,B) and the corresponding mean decision time (C,D)
for two different decision processes. (A,C) For the dKPR process with varying kinetic proofreading rates
$\psi_1= \gamma_1/k_1$ and $\psi_2= \gamma_2/k_2$ and zero backward rates, $r_{1,2}=0$.
(B,D)  For the AM process with varying backward rates $\theta_1= r_1/k_1$ and $\theta_2= r_2/k_2$ and zero
proofreading rates, $\gamma_{1,2}=0$. For both systems, we have set the forward rates to $k_1=6$, $k_2=1$,
and the lengths to $L_1=L_2 = 16$. The contour lines denote the probabilities of correct completion (upper panels)
and mean completion time in units of $1/k_2$ (lower panels). The red dashed line corresponds to a twenty percent
difference in the proofreading or backward {\em ratios}, $\psi_1=0.8\psi_2$ or $\theta_1=0.8\theta_2$, respectively.
The blue dashed line corresponds to a twenty percent difference in the proofreading or backward {\em rates},
$\gamma_1=0.8\gamma_2$ or $r_1=0.8r_2$, respectively.}
\label{fig:dKPR_Correct2}
\end{figure*}

Figs.\ \ref{fig:dKPR_Correct2}A-B show the completion probabilities for a case where the forward rates are different
from one branch to the next. While many qualitative trends of this case are similar to the previous case with equal
forward rates, the analysis becomes a little more complicated. First, the fact of different rates already provides a
certain amount of correction ($k_1/(k_1+k_2)=6/7$) before any additional effects of proofreading or backward rates.
In turn, the proofreading and backward rates can amplify this specificity much higher than in the previous case for
similar relative changes in parameters from one branch to the next. In this case, because the two branches have
different forward rates, one can consider small relative changes in the ratios ($\psi$ or $\theta$, red dashed lines)
or in the absolute rates ($\gamma$ or $r$, blue dashed lines).  
In the former case, with a twenty percent change in the ratios ($\psi_1=0.8\psi_2$ or $\theta_1 = 0.8\theta_2$),
either process can attain a 90\% specificity (white stars) but only the AM process is capable of providing 99\%
specificity (pink star) within the parameter range shown in the figure.  In the latter case, when the actual rates $\gamma$ or $r$ are only slightly varied from one branch to the other another (blue dashed lines) far greater specificity is achievable with either model.  Indeed, a high level of specificity is achievable in either process even when these rates are identical so long as the forward rates are different (not shown). 

\subsection{Average Completion Times}

In the perspective of kinetic proofreading, in addition to forming the correct product, a process must complete this construction in a timely manner. For example, the AM and dKPR schemes may make the same amplification of specificity,
but one may be able to do so faster than the other. While a detailed analysis of this tradeoff between specificity
and efficiency is left for future work, we begin to explore this aspect of the system by examining the mean completion
time. Although the expressions for the mean completion times are trivial to generate, they are cumbersome to write
in the general case. Therefore, in the interest of brevity, we provide explicit expressions only for the case of
directed kinetic proofreading, for which the mean ``correct" completion time is given by
\begin{align}
T_{C-{\rm dKPR}}&=-\frac{ 
\left( k_1/k_2 \right)  \left( 1+\psi_1 \right) \left[ 1-\left( 1+\psi_2 \right)^{L_2} \right]+\psi_2  \left[ \left(1-L_1 \right) \left( 1+\psi_2 \right) + \left( k_1/k_2 \right) L_2 \left( 1+ \psi_1 \right) \right]}{k_1\psi_2\left(1+\psi_1\right)\left[\left(1+\psi_1\right)^{L_1}\left(1+\psi_2\right)+\left(k_1/k_2\right)\left(1+\psi_1\right)
\left(1+\psi_2\right)^{L_2}\right]} \nonumber \\
&-\frac{
\left( k_1/k_2 \right) \left( 1+\psi_2 \right)^{L_2} \left[ 1+\psi_1 \left( 2+\psi_1 \right) \right]
\left[ 1-\left( 1+\psi_1 \right)^{L_1} \right]}{k_1\psi_1\left(1+\psi_1\right)^{L_1+1}\left[\left(1+\psi_1\right)^{L_1}\left(1+\psi_2\right)
+\left(k_1/k_2\right)\left(1+\psi_1\right)\left(1+\psi_2\right)^{L_2}\right]}. \label{fir-mom} 
\end{align} 
%
%
%
%
%\begin{align}
%T_{C-{\rm dKPR}}&=
%-\frac{ 
%\left( k_1/k_2 \right)  \left( 1+\psi_1 \right) \left[ 1-\left( 1+\psi_2 \right)^{L_2} \right]+\psi_2  \left[ \left(1-L_1 \right) \left( 1+\psi_2 \right) + \left( k_1/k_2 \right) L_2 \left( 1+ \psi_1 \right) \right]}
%{k_1\psi_2\left[\left(1+\psi_1\right)^{L_1+1}\left(1+\psi_2\right)+\left(k_1/k_2\right)\left(1+\psi_1\right)^2
%\left(1+\psi_2\right)^{L_2}\right]} \nonumber \\
%&-\frac{
%\left( k_1/k_2 \right) \left( 1+\psi_2 \right)^{L_2-1}
%\left[ \left(1+\psi_1\right)^2-\left( 1+\psi_1 \right)^{L_1+2}
% \right]}
% {k_1\psi_1\left[\left(1+\psi_1\right)^{2L_1+1}
%+\left(k_1/k_2\right)\left(1+\psi_1\right)^{L_1+2}\left(1+\psi_2\right)^{L_2-1}\right]}. \label{fir-mom} 
%\end{align} 
%
%
%
Similarly, we find the mean ``wrong" completion time
\begin{align}\label{fir-mom-2}
T_{W-{\rm dKPR}}&=-\frac{\left[1-\left(1+\psi_1\right)^{L_1}\right]\left(1+\psi_2\right)^{L_2}+\left(\psi_1/\psi_2\right)\left[1-\left(1+\psi_2\right)^{L_2}\right]\left(1+\psi_1\right)^{L_1}\left(1+\psi_2\right)}{k_2\psi_1\left[\left(1+\psi_1\right)^{L_1}\left(1+\psi_2\right)+\left(k_1/k_2\right)\left(1+\psi_1\right)
\left(1+\psi_2\right)^{L_2}\right]} \nonumber \\
&-\frac{L_1\left(1+\psi_2\right)^{L_2}-\left(k_1/k_2\right)\left(L_2-1\right)\left(1+\psi_1\right)\left(1+\psi_2\right)^{L_2-1}}{k_2\left[\left(1+\psi_1\right)^{L_1}\left(1+\psi_2\right)+\left(k_1/k_2\right)\left(1+\psi_1\right)
\left(1+\psi_2\right)^{L_2}\right]}.
\end{align} 
The average arbitrary completion time (without specifying correct or wrong completion) is
\begin{align}\label{fir-mom-arb}
%\leftexp{dKPR}{T}&
{T_{\rm dKPR}}&
=\frac{\left(1+\psi_1\right)^{L_1}\left(1+\psi_2\right)^{L_2}-\left(1+\psi_1\right)^{L_1}-\left(1+\psi_2\right)^{L_2}}{k_2\left[\left(1+\psi_1\right)^{L_1}\left(1+\psi_2\right)+\left(k_1/k_2\right)\left(1+\psi_1\right)
\left(1+\psi_2\right)^{L_2}\right]} \nonumber \\
&-\frac{\left(1+\psi_1\right)^{L_1}\left[1-\left(1+\psi_2\right)^{L_2}\right]+\left(\psi_2/\psi_1\right)\left(1+\psi_1\right)^{L_2}\left[1-\left(1+\psi_1\right)^{L_1}\right]}{k_2\psi_2\left[\left(1+\psi_1\right)^{L_1}\left(1+\psi_2\right)+\left(k_1/k_2\right)\left(1+\psi_1\right)
\left(1+\psi_2\right)^{L_2}\right]}.
\end{align}

Figs.\ \ref{fig:dKPR_Correct}C-D show contour plots for the average completion times of the dKPR and AM processes
for ranges comparable to the specificity plots in Figs.\ \ref{fig:dKPR_Correct}A-B and $k_{1,2}=1$.
From these plots, we can observe that as the backward or proofreading rates increase, the amount of time required
to complete the process increases exponentially. As before in Figs.\ \ref{fig:dKPR_Correct}A-B, the dashed line denotes
the lines where $\psi_1=0.8\psi_2$ or $\theta_1=0.8\theta_2$ and the stars represent the crossings of the 90\%
specificity. 
While we saw in Figs.\ \ref{fig:dKPR_Correct}A-B that both processes were able to provide 90\% specificity (for 20\% difference in the backward/proofreading rates), the AM process can provide it with a much smaller mean completion time.
Similarly, Figs.\ \ref{fig:dKPR_Correct2}C-D show contour plots of the mean completion times of the dKPR and AM
processes with $k_1=1$ and $k_2=6$.
The white/pink/black stars denote the 90\%, 99\%, 99.9\% specificities correspondingly.
The red dashed lines correspond to $\theta_1=0.8\theta_2$ (or $\psi_1=0.8\psi_2$) and the blue dashed lines correspond
to $r_1=0.8r_2$ (or $\gamma_1=0.8\gamma_2$). We can see again that for a 20\% difference backward/proofreading rates (blue dashed lines) or their ratios to the corresponding forward rates (the red dashed lines) the AM process can
provide the requested specificity for much smaller average completion times.

\begin{figure*}[t]
\includegraphics[width=0.85\linewidth]{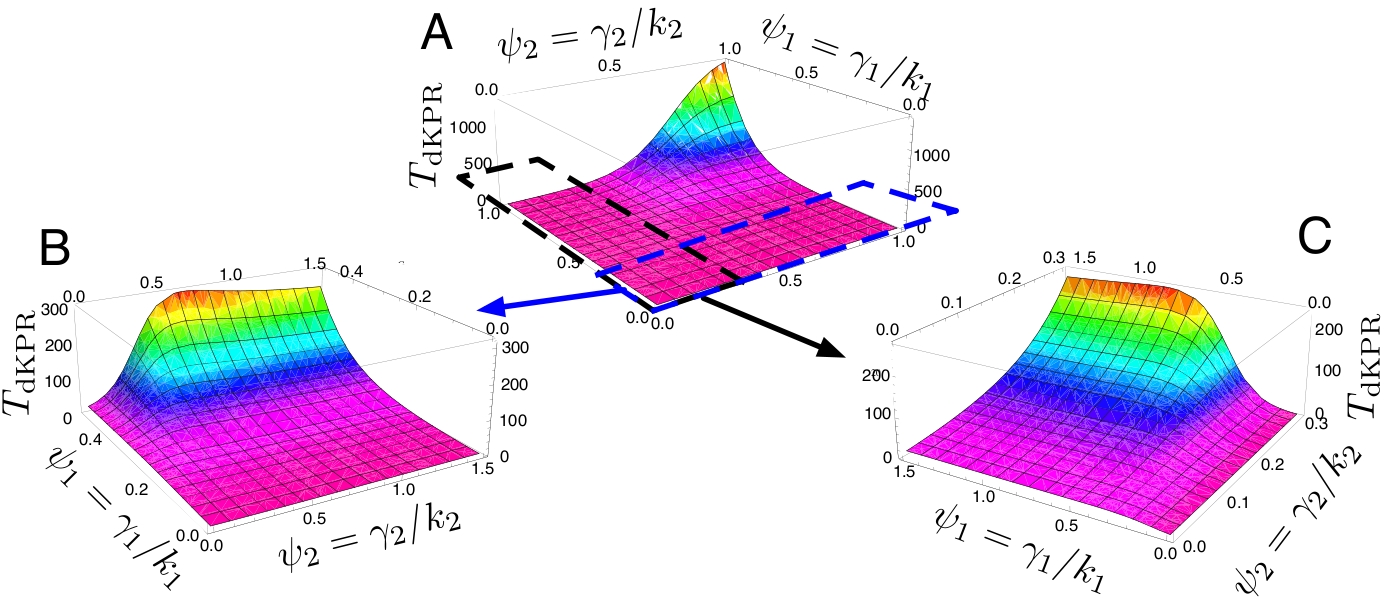}
\caption{
Plots of the mean arbitrary completion times (units of $1/k_2$) for the directed kinetic proofreading process, with two branches
of lengths $L_1=L_2=16$, forward rates $k_1=6$ and $k_2=1$. Panels B,C show a zoomed in perspective of the mean
completion times corresponding to the parameter regions indicated in panel A.}
\label{fig:Mean_dKPR}
\end{figure*}

To better understand the behavior of the mean completion time, we illustrate in Fig.\ \ref{fig:Mean_dKPR} the effects
that changes in the parameters $\psi_{1,2}$ have on these mean completions times for the process in which the forward
rate on the correct branch is six times the rate on the wrong branch, $k_1=6k_2$. 
At first glance at Fig.\ \ref{fig:Mean_dKPR}A or Fig.\ \ref{fig:dKPR_Correct2}C it appears that the behavior of the mean arbitrary completion time is somewhat trivial,
as one increases the proofreading rates in both branches, the mean waiting time also increases.
However, by zooming in along certain strips of this plot, one finds additional dependencies of the mean waiting times
on the parameters. Suppose that one fixes $\psi_{1}$ to some non-zero value and then changes $\psi_2$ (see top edge of
Fig.\ \ref{fig:Mean_dKPR}B). When $\psi_2$ is zero, the second branch is biased forward and the process will quickly
complete soon after it enters into that branch. Conversely, when $\psi_2$ is very large, the process will spend very
little time in the second branch and the process reduces down to the single branch process as if that second branch were
not there. However, when $\psi_2$ is in some middle range, the process will spend significant amounts of time in each of
the two branches, thereby increasing the total time until the completion. Similar observations can be made for the AM
process (not shown), as should be expected from the non trivial shape of the contours of Fig.\ \ref{fig:dKPR_Correct2}D.
  
\subsection{Variance in Completion Times}
In addition to specificity and the average time to arrive at that specificity, a completion process is further characterized by the
shape of the distribution for its completion time. For some parameters this process will have little variance, and the
decision is made in some seemingly fixed deterministic amount of time. For other parameters this decision may be much
more broadly distributed (the same behavior was found for a single branch processes, see \cite{BMN:2009}).
The relative broadness of this shape can be described by the squared coefficient of variation (variance divided by the mean
squared, ${\rm CV}^2=\sigma^2/\mu^2$) of the completion time distribution.
The second moments, and therefore the variances, can be derived according to the general relation of 
Eqs.~(\ref{align},\ref{moments2}), but the resulting expressions are too long to provide much valuable insight
even in the case of directed kinetic proofreading. Instead, we rely on parametric studies to explore how parameters
affect the completion time distribution shapes.

\begin{figure*}[t]
\includegraphics[width=\linewidth]{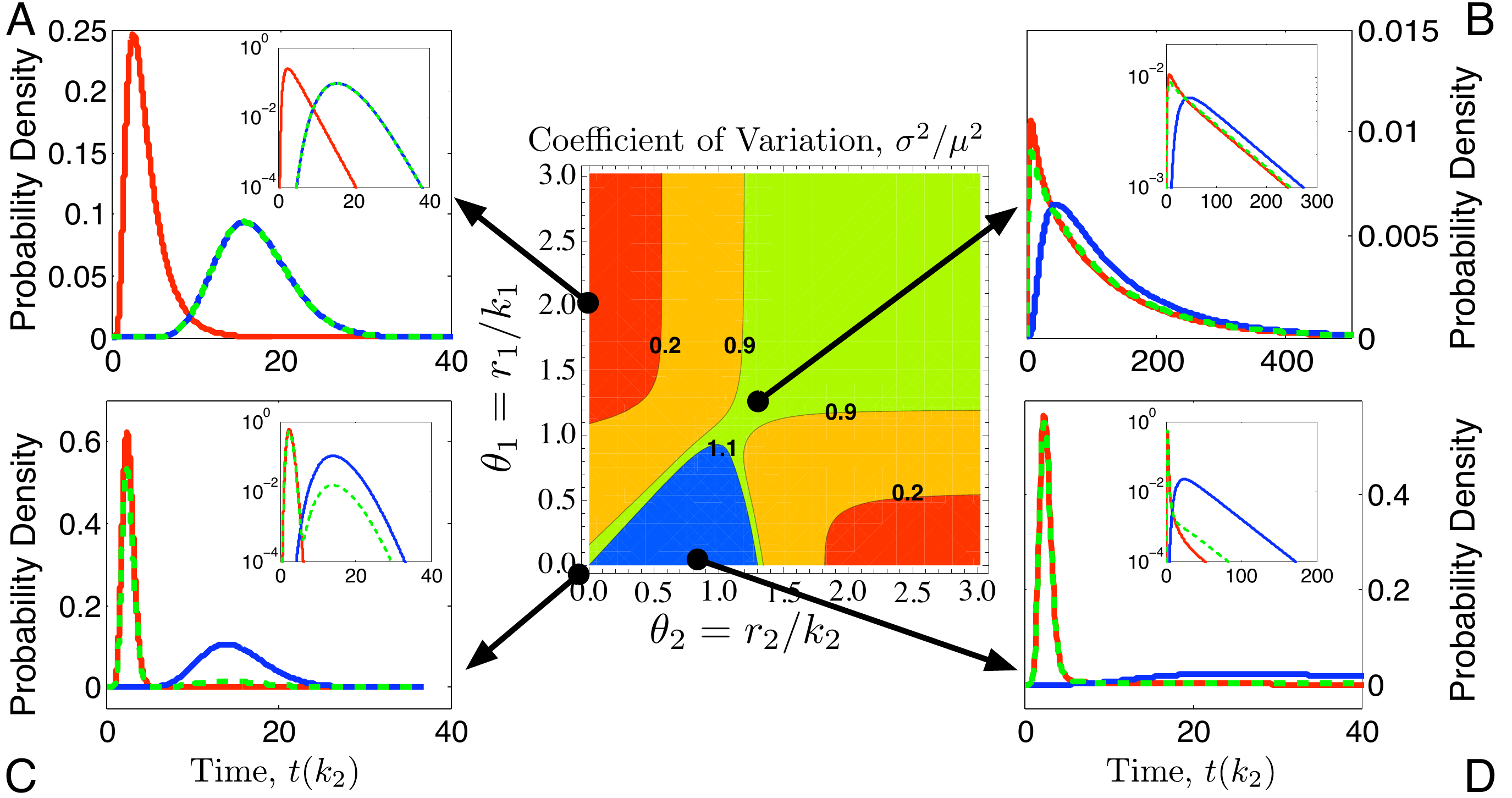}
\caption{
Contour plot of the coefficient of variation (of the arbitrary completion time) versus $r_1/k_1$ and $r_2/k_2$ and typical completion time distributions.
We used the case of zero proofreading rates, $\gamma_{1,2}=0$. We also set $k_1=6$ and $k_2=1$. The different colors 
correspond to different behavior of the completion time distributions (see text for more details). 
The side panels (A-D) show the distributions of completion times in the correct (red) and incorrect (blue) directions 
and the arbitrary completion time distribution (green). The inset in each of the panels shows a semi log plot of the 
distribution to amplify the differences between the lines.}
\label{fig:2}
\end{figure*}

In what follows we consider the same cases as above and classify the shapes of the resulting completion time
distributions. First, we consider the case of zero proofreading rates, $\gamma_{1,2}=0$.
Fig.~\ref{fig:2} shows a contour plot of the coefficient of variation of the arbitrary completion time versus $\theta_1=r_1/k_1$ and
$\theta_{2}= r_2/k_2$ and typical completion time distributions for the parameter values $k_1=6k_2$ and
$\{(\theta_1,\theta_2)\}$ = $\{(2,1),\ (1.2,1.2),\ (0,0),\ (0,0.88)\}$.
This plot allows us to divide the parameters space into a few regions with different shapes for the completion time
distribution. The large green area (color online) in the upper right corner corresponds to $0.9<{\rm CV}^2<1.1$,
where the completion time distribution is often well approximated by an exponential distribution.
The corresponding side panel (Fig.\ \ref{fig:2}B) shows the ``correct" (red) and ``wrong" (blue) completion time
distributions as well as the arbitrary completion time distribution (green). In this case, all three distributions
are almost exponential, with the small exception of their left tails. The red areas (color online) where the coefficient
of variation is $0<{\rm CV}^2<0.2$ correspond to cases where one branch is strongly biased backwards while the other is
biased forward. For these, the completion time along the backward biased branch is nearly exponential, while the
completion time along the forward biased branch is effectively described by a gamma distribution 
(see Fig.\ \ref{fig:2}A). Since the process is far more likely to finish along the forward biased branch, the total
completion time distribution is also well approximated by a narrow Gamma distribution as illustrated in 
Fig.\ \ref{fig:2}A. The bottom left panel shows the distributions in the case where both branches are biased forward
$r_1=r_2=0$. In this case, the completion time distribution for each branch is a Gamma distribution, and the total
completion time distribution is a simple combination of the two, since the probability to complete at each of the
branches is proportional to the forward rate at that branch. As a result the total completion time has a bimodal 
distribution as shown in Fig.\ \ref{fig:2}C. The final area of interest (shown in blue online) corresponds to the 
conditions where the coefficient of variation is greater than $1.1$, such that the total completion time distribution
is broader than exponential as is shown in Fig.\ \ref{fig:2}D for the point of maximal ${\rm CV}^2$.
Due to the fact that motion in one branch is strongly biased forward while motion in the other branch is almost 
unbiased, we obtain a non-trivial combination of the two behaviors in the total completion time distribution.

\begin{figure*}[t]
\includegraphics[width=\linewidth]{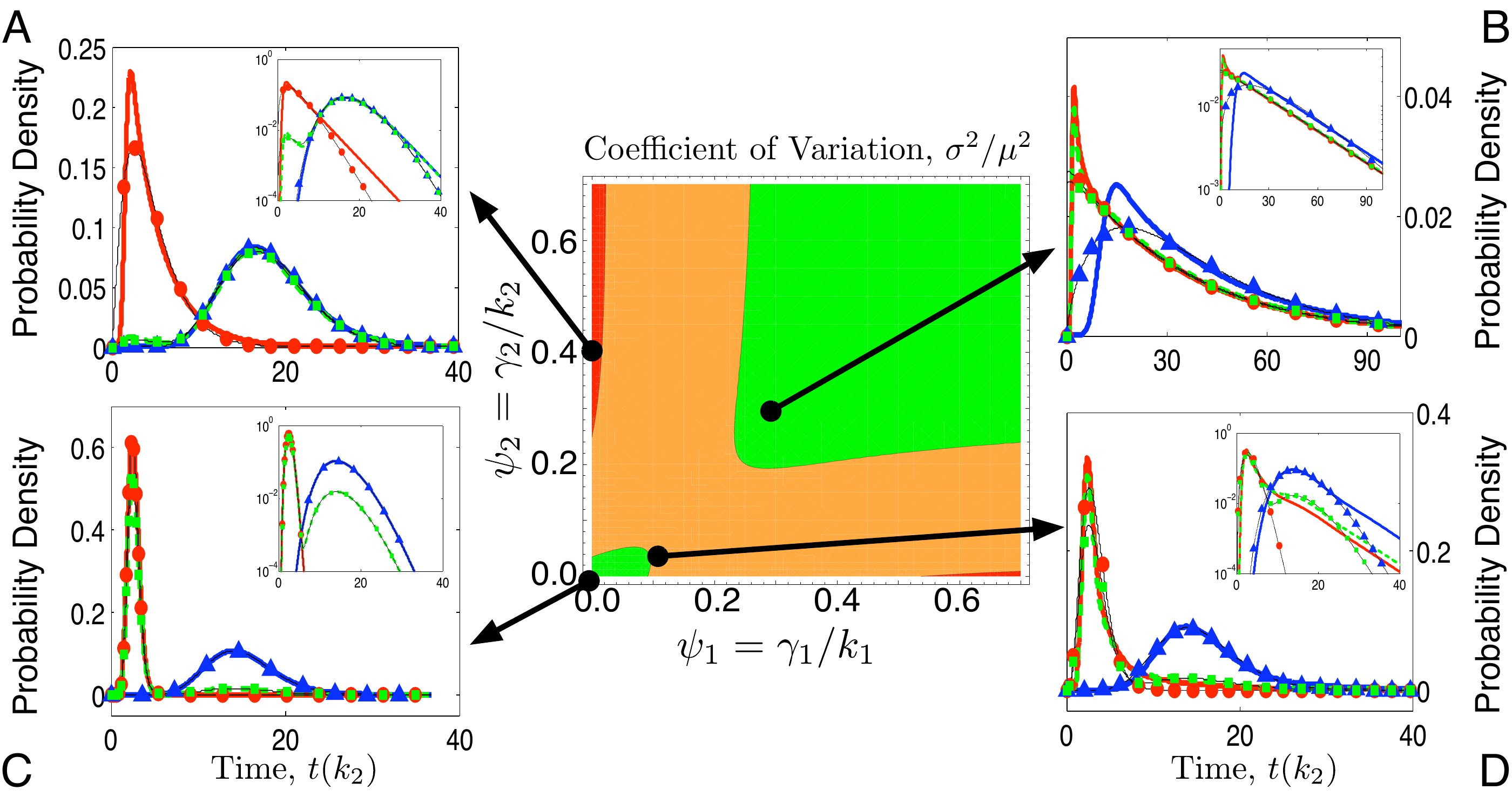}
\caption{
Contour plot of the coefficient of variation (of the arbitrary completion time) versus $\psi_1=\gamma_1/k_1$ and $\psi_2=\gamma_2/k_2$ and typical 
completion time distributions. We used the case of zero backward rates, $r_{1,2}=0$. We also set $k_1=1$ and $k_2=6$. 
The different colors correspond to different behavior of the completion time distributions (see text for more details). 
The side panels show the distributions of completion times in the correct (blue) and incorrect (red) directions and the 
arbitrary completion time distribution (green). The markers correspond to the best fit for a reduced 3-state model 
approximation to the processes. The inset in each of the panels shows a semi log plot of the distribution to amplify 
the differences between the lines.}
\label{fig:dKPR_Contours}
\end{figure*}

We now consider the case where there is proofreading ($\gamma_{1,2} >0$) but where the backward rates are set to zero,
$r_{1,2}=0$. Fig.~\ref{fig:dKPR_Contours} shows a contour plot of the coefficient of variation of the arbitrary completion time versus 
$\psi_1=\gamma_1/k_1$ and $\psi_{2}= \gamma_2/k_2$ and typical completion time distributions for the parameter
values $k_1=6k_2$ and $\{(\psi_1,\psi_2)\}$ = $\{(0.4,0),\ (0.3,0.3),\ (0,0),\ (0.05,0.1)\}$.  
As above in Fig.\ \ref{fig:2}, we can divide the parameters space into few regions with different shapes for the 
completion time distribution. For example the large green area (color online) corresponds to a coefficient of variation
near one and where the directional and arbitrary completion time distributions are well approximated by exponential 
distributions (see Fig.\ \ref{fig:dKPR_Contours}B). Similarly, for the small red areas where one branch is biased 
backwards and the other forward, the completion time along the backward biased branch is nearly exponential, 
while the completion time along the forward biased branch is effectively described by a gamma distribution 
(see Fig.\ \ref{fig:dKPR_Contours}A).  

\begin{figure*}
\includegraphics[width=\linewidth]{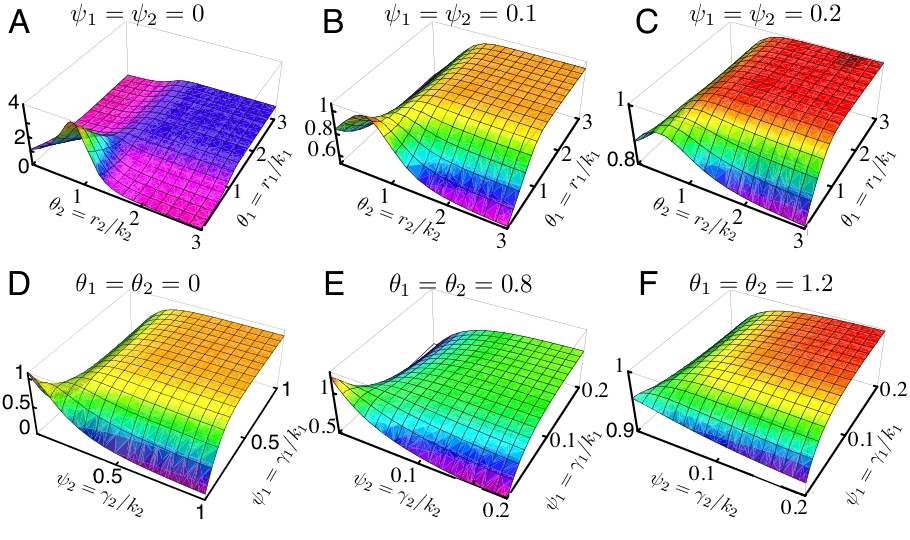}
\caption{
The coefficient of variation versus $\theta_1$ and $\theta_2$ or $\psi_1$ and $\psi_2$. 
In the upper line we fix the ratio between the proofreading rate and the forward rate ($\psi_{1,2}$) in both branches 
and show the effect of changing the ratios between the backward and forward rates $\theta_{1,2}$. In the bottom line we 
fix $\theta_{1,2}$ and show the effect of changing $\psi_{1,2}$. In all cases as both branches are strongly backward 
biased $CV\sim1$ and the completion time distribution is exponential. Further discussion appears in the text.}
\label{fig:3}
\end{figure*}

We now turn to the more general case where there is both proofreading and a backward reactions 
($\gamma_{1,2} >0$, $r_{1,2}>0$). For this case, Fig.~\ref{fig:3} shows a 3D plot of the coefficient of variation of the arbitrary completion time vs. 
$\theta_{1,2}$ (upper line) or $\psi_{1,2}$ lower line. These figures emphasize the different effect of changes in 
$\theta$ or $\psi$. While in all cases strong backward bias on both branches (large $\theta_{1,2}$ or $\psi_{1,2}$) 
lead to an exponential distribution of the completion time, backward bias has different dependence on the system size 
and different ranges for $\theta$ and $\psi$.  

\subsection{Simplification of the Two-Branch Decision Process}

\begin{figure*}[t]
\includegraphics[width=4in]{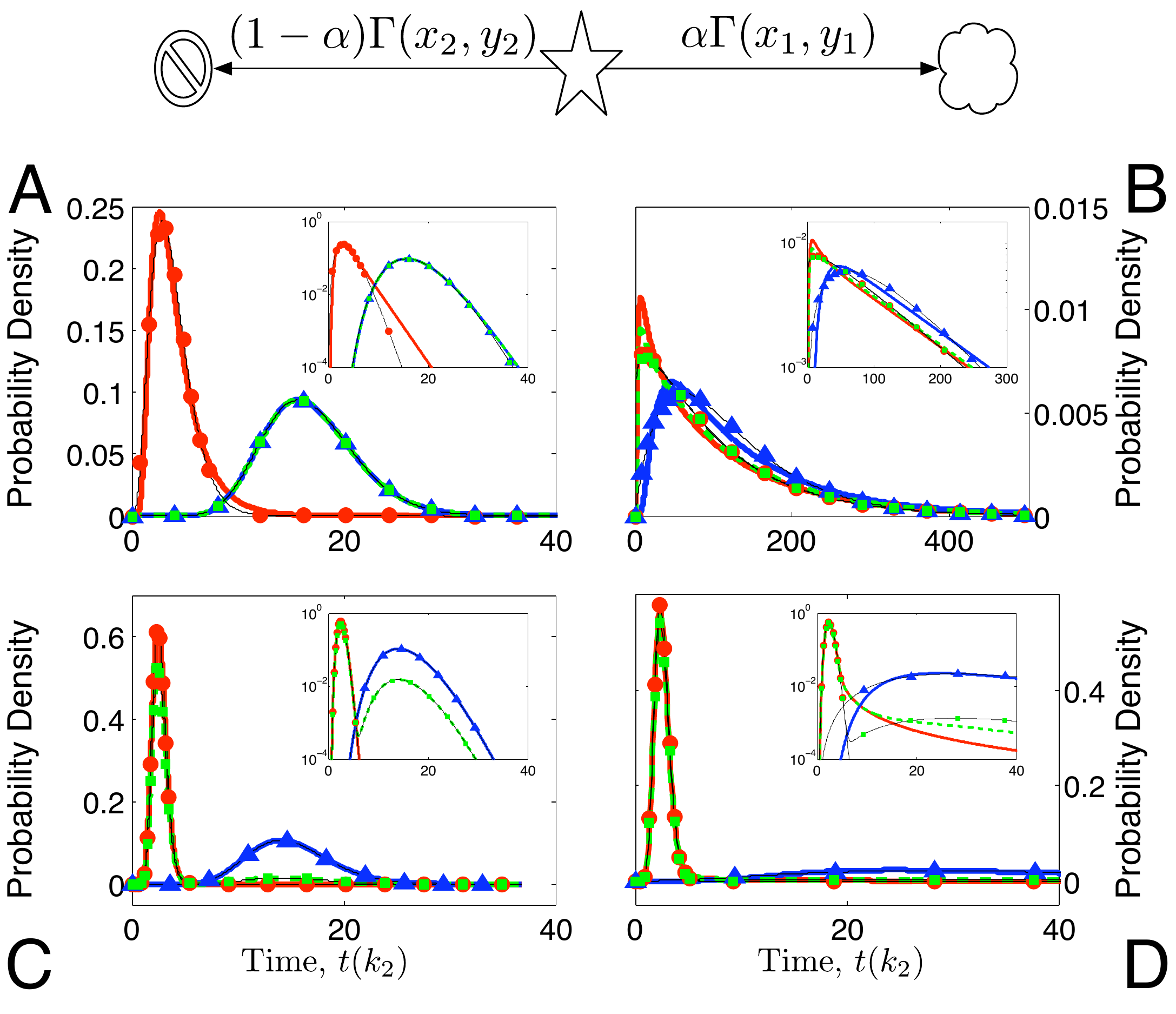}
\caption{
Three state model approximation of the original completion time problem.  
(Top) Schematic description of the three state model where the conditional escape time in each direction is given by a gamma 
distribution. (A-D) Comparison of the escape time distributions using the full original and the reduced three state 
model. The parameters used here are the same as those in Figs.\ \ref{fig:2}(A-D).}
\label{fig:2b}
\end{figure*}

\begin{figure*}[t]
\includegraphics[width=\linewidth]{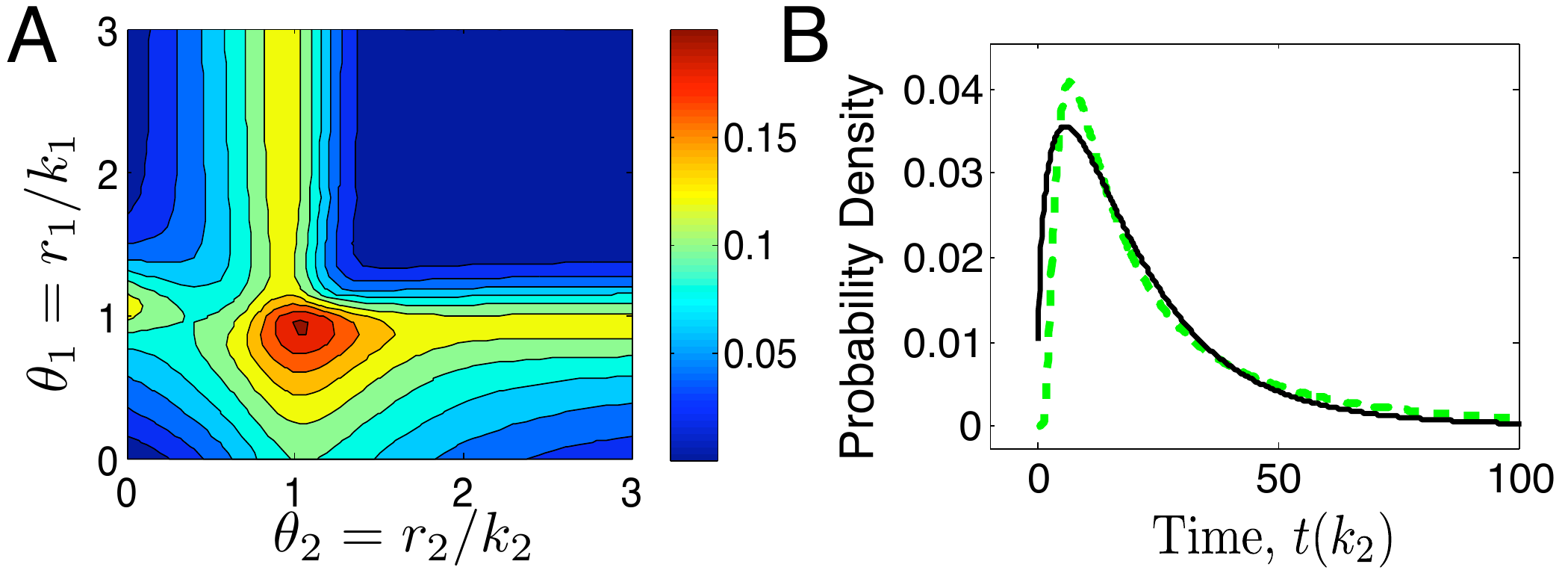}
\caption{
Numerical comparison of the completion time distributions for the approximate 3-state model and the full two branch 
process. (A) Contour plots of the approximation error (the norm of the difference between the actual and the approximate 
joint distributions (see Eq.~(\ref{fit}))) versus the ratios ($\theta_{1,2} = r_{1,2}/k_{1,2}$). 
(B) Illustration of approximate (dashed line) and actual (solid line) completion time distributions (in units of $1/k_2$) 
for the parameter set ($\theta_1=1.03$, $\theta_2=0.95$), which corresponds to the {\em largest} approximation error of 
$\sum_{n=1}^2 \int_{0}^{\infty} \left| f_n(t) - \tilde{f}_n(t) \right|_1 dt = 0.20$.}
\label{fig:Objective}
\end{figure*}

In examining the distributions in Figs.\ \ref{fig:2}A-D, one observes that the completion time distribution of each 
branch is often similar to a gamma distribution (or an exponential distribution, which is a special case of the gamma 
distribution). This suggests that one should frequently be able to replace the entire process with a simple three state 
chain as shown in Fig.\ \ref{fig:2b} with the following properties. Each direction (1,2) is assumed to have a 
non-normalized Gamma distributed completion time with density
\begin{align*}
f_1(t) &\approx \tilde{f}_1(t,x_1,y_1) = \alpha t^{x_1-1}y_1^{x_1}\frac{\exp(-y_1 t)}{\Gamma(x_1)}, \\
f_2(t) &\approx \tilde{f}_2(t,x_2,y_2) = (1-\alpha)t^{x_2-1}y_2^{x_2}\frac{\exp(-y_2 t)}{\Gamma(x_2)},
\end{align*}
where $0\leq \alpha \leq 1$ denotes the probability of completion in the first direction. 
Thus, the total probability density of completing along either branch at time $t$ is approximated by:
\[
f_T(t) \approx \tilde{f}_T(t) = \tilde{f}_1(t,x_1,y_1) + \tilde{f}_2(t,x_2,y_2).
\]
In numerical studies, we have attempted to find parameter sets ${\bf \Lambda} =\{x_1,y_1,x_2,y_2,\alpha\}$ 
that best match the direction and time distribution of the full escape process in the one norm sense.  
In other words, we have found the ${\bf \Lambda}$ such that:
\begin{equation}
{\bf \Lambda} =\text{arg}\hspace{-22pt}\min_{\{x_1,y_1,x_2,y_2,\alpha\}} \sum_{n=1}^2 \int_{0}^{\infty} \left|f_n(t) - \tilde{f}_n(t,\Lambda) \right|_1dt.\label{fit}
\end{equation}
In most cases, we find that this approximation and optimization does an excellent job of capturing the qualitative and 
quantitative behaviors of the complete process as is shown in Figs. \ref{fig:2b}A-D.  
To further explore the ability of the reduced model to capture the behavior of the full system, we have explored the 
original parameter space $\{\theta_1,\theta_2\}$ in order to find the regions where this approximation is most valid.  
From Fig. \ref{fig:Objective}A, we immediately see that the approximation is valid in all four corners of the contour 
plot where both $\theta_1$ and $\theta_2$ are either relatively large or relatively small--that is where both branches 
are biased in one direction or another. However, even in the regions where one or both branches are unbiased 
($\theta_1\approx 1$ or $\theta_2\approx 1$), we note that the fit is still quite good.  
Indeed for this system, we can always find a parameter set $\{x_1,y_1,x_2,y_2,\alpha\}$ that captures the full escape 
time distribution within error (defined by the norm in Eq.~(\ref{fit})) of 0.2. In order to illustrate this approximation success, 
Fig. \ref{fig:Objective}B shows the actual (solid line) and approximate (dashed line) distributions for the case 
($\theta_1=1.03$, $\theta_2=0.95$), of the {\em worst} fit. For every other case, we were able to find a three state 
model that did an even better job of matching the full system behavior.

As was the case for the AM process ($\gamma_{1,2}=0$), the dKPR process ($r_{1,2}=0$) is well captured by the same three state process defined above. To illustrate this, 
the colored lines in Figs.\ \ref{fig:dKPR_Contours}A-D correspond to the full system completion time distributions, and 
the markers correspond to the approximate three state system.

%____________________________________________
%____________________________________________

\section{Conclusions}
In this work we have begun the exploration of the temporal properties of kinetic proofreading schemes.
To accomplish this, we have derived analytical expressions for the Laplace transform of the occupation probabilities 
from which we obtained the completion time distributions.  
With this analysis, we have enabled the simple derivation of expressions for the completion time moments.  
Some of these expressions, such as completion probabilities and the mean waiting times for certain processes are 
simple enough to be shown explicitly, while others are just as easily derived, but are omitted since their form is too 
long and not very informative. To enable a better understanding of the interplay of specificity and temporal behaviors, 
we focused on the first two moments of the completion times as well as on the completion probabilities 
(which is actually the zeroth moment).
We showed that for most parameter sets, each of the considered proofreading schemes can be reduced to a three 
state process with simple distributions for the waiting times between transitions. The simplified process captures most of the relevant features of kinetic proofreading schemes, namely, the specificity as well as the magnitude and shape of the completion time distributions.  However, the dependence of the simplified behavior on the full system's kinetic parameters is different for the various proofreading schemes, suggesting that some important information about the process is retained despite the simplification.

We have explicitly considered different kinetic schemes including the 
traditional directed kinetic proofreading (dKPR) scheme where catastrophic reactions force the process to restart 
as well as an absorption mode (AM) where single step intermediate reactions can provide the same specificity.  
Surprisingly, we find that in most cases the simpler AM process outperforms the dKPR process by providing a higher 
degree of specificity in a shorter amount of time. It is also worth mentioning that the dKPR or 
general kinetic proofreading processes violate the detailed balance conditions and therefore are necessarily 
non-equilibrium processes. The AM process on the other hand may satisfy the detailed balance condition and in 
this case is an equilibrium process.  In this sense, the AM process has the added advantage in that it conserves energy, while the dKPR process must be continually driven with externally applied energy.

High specificity appears in many biological systems and likely results from many different kinetic schemes--suggesting that one needs as much information as possible to distinguish between one such mechanism and the next.
Therefore, in addition to using the specificity and mean completion times to compare the different processes, 
we have also used analyses of the completion time distributions to classify different kinetic schemes and parameter values into separate regimes where these distributions take on different qualitative shapes. 
By providing this additional information, the temporal analysis and classification tools developed here can more precisely support or oppose hypotheses of particular kinetic proofreading models for particular biochemical systems.  In the future, the next logical step is to apply these tools in order to identify parameters and infer kinetic mechanisms from experimental measurements of completion time distributions.

\begin{acknowledgments}
We thank N. Hengartner for discussions during early stages of
  this work. We also thank B.\ Goldstein, R.\ Gutenkunst, M.\
  Monine, and M. Savageau for helpful comments regarding this work. This work was partially funded by LANL LDRD program.

\end{acknowledgments}
\bibliography{KPR-PRE-2}

\begin{thebibliography}{37}
\expandafter\ifx\csname natexlab\endcsname\relax\def\natexlab#1{#1}\fi
\expandafter\ifx\csname bibnamefont\endcsname\relax
  \def\bibnamefont#1{#1}\fi
\expandafter\ifx\csname bibfnamefont\endcsname\relax
  \def\bibfnamefont#1{#1}\fi
\expandafter\ifx\csname citenamefont\endcsname\relax
  \def\citenamefont#1{#1}\fi
\expandafter\ifx\csname url\endcsname\relax
  \def\url#1{\texttt{#1}}\fi
\expandafter\ifx\csname urlprefix\endcsname\relax\def\urlprefix{URL }\fi
\providecommand{\bibinfo}[2]{#2}
\providecommand{\eprint}[2][]{\url{#2}}

\bibitem[{\citenamefont{Hopfield}(1974)}]{Hopfield:1974}
\bibinfo{author}{\bibfnamefont{J.}~\bibnamefont{Hopfield}},
  \bibinfo{journal}{Proc. Natl. Acad. Sci. (USA)}
  \textbf{\bibinfo{volume}{71}}, \bibinfo{pages}{4135} (\bibinfo{year}{1974}).

\bibitem[{\citenamefont{Yan et~al.}(1999)\citenamefont{Yan, Magnasco, and
  Marko}}]{Yan:1999}
\bibinfo{author}{\bibfnamefont{J.}~\bibnamefont{Yan}},
  \bibinfo{author}{\bibfnamefont{M.}~\bibnamefont{Magnasco}}, \bibnamefont{and}
  \bibinfo{author}{\bibfnamefont{J.}~\bibnamefont{Marko}},
  \bibinfo{journal}{Nature} \textbf{\bibinfo{volume}{401}},
  \bibinfo{pages}{932} (\bibinfo{year}{1999}).

\bibitem[{\citenamefont{Sancar et~al.}(2004)\citenamefont{Sancar, Unsal-Kacmaz,
  and Linn}}]{Sancar:2004}
\bibinfo{author}{\bibfnamefont{A.}~\bibnamefont{Sancar}},
  \bibinfo{author}{\bibfnamefont{K.}~\bibnamefont{Unsal-Kacmaz}},
  \bibnamefont{and} \bibinfo{author}{\bibfnamefont{S.}~\bibnamefont{Linn}},
  \bibinfo{journal}{Ann. Rev. Biochem.} \textbf{\bibinfo{volume}{73}},
  \bibinfo{pages}{39} (\bibinfo{year}{2004}).

\bibitem[{\citenamefont{Goulian et~al.}(1968)\citenamefont{Goulian, Lucas, and
  Kronberg}}]{Goulian}
\bibinfo{author}{\bibfnamefont{M.}~\bibnamefont{Goulian}},
  \bibinfo{author}{\bibfnamefont{Z.~J.} \bibnamefont{Lucas}}, \bibnamefont{and}
  \bibinfo{author}{\bibfnamefont{A.}~\bibnamefont{Kronberg}},
  \bibinfo{journal}{J.\ Biol.\ Chem.} \textbf{\bibinfo{volume}{243}},
  \bibinfo{pages}{627} (\bibinfo{year}{1968}).

\bibitem[{\citenamefont{Blanchard et~al.}(2004)}]{Blanchard:2004}
\bibinfo{author}{\bibfnamefont{S.}~\bibnamefont{Blanchard}}
  \bibnamefont{et~al.}, \bibinfo{journal}{Nature Struct. Mol. Biol.}
  \textbf{\bibinfo{volume}{11}}, \bibinfo{pages}{1008} (\bibinfo{year}{2004}).

\bibitem[{\citenamefont{Jovanovic-Talisman et~al.}(2008)}]{NPC}
\bibinfo{author}{\bibfnamefont{T.}~\bibnamefont{Jovanovic-Talisman}}
  \bibnamefont{et~al.}, \bibinfo{journal}{Nature}
  \textbf{\bibinfo{volume}{457}}, \bibinfo{pages}{1023} (\bibinfo{year}{2008}).

\bibitem[{\citenamefont{Mckeitan}(1995)}]{McKeithan:1995}
\bibinfo{author}{\bibfnamefont{T.}~\bibnamefont{Mckeitan}},
  \bibinfo{journal}{Proc. Natl. Acad. Sci. (USA)}
  \textbf{\bibinfo{volume}{92}}, \bibinfo{pages}{5042} (\bibinfo{year}{1995}).

\bibitem[{\citenamefont{Rabinowitz et~al.}(1996)}]{Rabinowitz:1996}
\bibinfo{author}{\bibfnamefont{J.}~\bibnamefont{Rabinowitz}}
  \bibnamefont{et~al.}, \bibinfo{journal}{Proc. Natl. Acad. Sci. (USA)}
  \textbf{\bibinfo{volume}{93}}, \bibinfo{pages}{1401} (\bibinfo{year}{1996}).

\bibitem[{\citenamefont{Rosette et~al.}(2001)}]{Rosette:2001}
\bibinfo{author}{\bibfnamefont{C.}~\bibnamefont{Rosette}} \bibnamefont{et~al.},
  \bibinfo{journal}{Immunity} \textbf{\bibinfo{volume}{15}},
  \bibinfo{pages}{59} (\bibinfo{year}{2001}).

\bibitem[{\citenamefont{Liu et~al.}(2001)}]{Liu:2001}
\bibinfo{author}{\bibfnamefont{Z.}~\bibnamefont{Liu}} \bibnamefont{et~al.},
  \bibinfo{journal}{Proc. Natl. Acad. Sci. (USA)}
  \textbf{\bibinfo{volume}{98}}, \bibinfo{pages}{7289} (\bibinfo{year}{2001}).

\bibitem[{\citenamefont{Goldstein et~al.}(2004)\citenamefont{Goldstein, Faeder,
  and Hlavacek}}]{Goldstein:2004}
\bibinfo{author}{\bibfnamefont{B.}~\bibnamefont{Goldstein}},
  \bibinfo{author}{\bibfnamefont{J.}~\bibnamefont{Faeder}}, \bibnamefont{and}
  \bibinfo{author}{\bibfnamefont{W.}~\bibnamefont{Hlavacek}},
  \bibinfo{journal}{Nature Rev. Immunol.} \textbf{\bibinfo{volume}{4}},
  \bibinfo{pages}{445} (\bibinfo{year}{2004}).

\bibitem[{\citenamefont{Faeder et~al.}(2003)}]{faeder-03}
\bibinfo{author}{\bibfnamefont{J.}~\bibnamefont{Faeder}} \bibnamefont{et~al.},
  \bibinfo{journal}{J. Immunol.} \textbf{\bibinfo{volume}{170}},
  \bibinfo{pages}{3769} (\bibinfo{year}{2003}).

\bibitem[{\citenamefont{Springgate and Loeb}(1975)}]{Springgate}
\bibinfo{author}{\bibfnamefont{C.~F.} \bibnamefont{Springgate}}
  \bibnamefont{and} \bibinfo{author}{\bibfnamefont{L.~A.} \bibnamefont{Loeb}},
  \bibinfo{journal}{J. Mol. Biol.} \textbf{\bibinfo{volume}{97}},
  \bibinfo{pages}{577} (\bibinfo{year}{1975}).

\bibitem[{\citenamefont{Ninio}(1975)}]{Ninio-1}
\bibinfo{author}{\bibfnamefont{J.}~\bibnamefont{Ninio}},
  \bibinfo{journal}{Biocheimie} \textbf{\bibinfo{volume}{57}},
  \bibinfo{pages}{587} (\bibinfo{year}{1975}).

\bibitem[{\citenamefont{Freter and Savageau}(1980)}]{Freter}
\bibinfo{author}{\bibfnamefont{R.~R.} \bibnamefont{Freter}} \bibnamefont{and}
  \bibinfo{author}{\bibfnamefont{M.}~\bibnamefont{Savageau}},
  \bibinfo{journal}{J.\ Theor.\ Biol.} \textbf{\bibinfo{volume}{85}},
  \bibinfo{pages}{99} (\bibinfo{year}{1980}).

\bibitem[{\citenamefont{Savageau}(1981)}]{Savageau}
\bibinfo{author}{\bibfnamefont{M.}~\bibnamefont{Savageau}},
  \bibinfo{journal}{J.\ Theor.\ Biol.} \textbf{\bibinfo{volume}{93}},
  \bibinfo{pages}{179} (\bibinfo{year}{1981}).

\bibitem[{\citenamefont{Zilman et~al.}()\citenamefont{Zilman, Pearson, and
  Bel}}]{AZJPGB}
\bibinfo{author}{\bibfnamefont{A.}~\bibnamefont{Zilman}},
  \bibinfo{author}{\bibfnamefont{J.}~\bibnamefont{Pearson}}, \bibnamefont{and}
  \bibinfo{author}{\bibfnamefont{G.}~\bibnamefont{Bel}},
  \eprint{cond-mat/0907.3160}.

\bibitem[{\citenamefont{D'Orsogna and Chou}(2005)}]{Chou:2005}
\bibinfo{author}{\bibfnamefont{M.}~\bibnamefont{D'Orsogna}} \bibnamefont{and}
  \bibinfo{author}{\bibfnamefont{T.}~\bibnamefont{Chou}},
  \bibinfo{journal}{Phys. Rev. Lett.} \textbf{\bibinfo{volume}{95}},
  \bibinfo{pages}{170603} (\bibinfo{year}{2005}).

\bibitem[{\citenamefont{Redner}(2001)}]{redner:2001}
\bibinfo{author}{\bibfnamefont{S.}~\bibnamefont{Redner}},
  \emph{\bibinfo{title}{A Guide To First-Passage Processes}}
  (\bibinfo{publisher}{Cambridge University Press}, \bibinfo{year}{2001}).

\bibitem[{\citenamefont{Bel and Barkai}(2005)}]{Bel:2005PRL}
\bibinfo{author}{\bibfnamefont{G.}~\bibnamefont{Bel}} \bibnamefont{and}
  \bibinfo{author}{\bibfnamefont{E.}~\bibnamefont{Barkai}},
  \bibinfo{journal}{Phys. Rev. Lett.} \textbf{\bibinfo{volume}{94}},
  \bibinfo{pages}{240602} (\bibinfo{year}{2005}).

\bibitem[{\citenamefont{Bel and Barkai}(2006)}]{Bel:2006PRE}
\bibinfo{author}{\bibfnamefont{G.}~\bibnamefont{Bel}} \bibnamefont{and}
  \bibinfo{author}{\bibfnamefont{E.}~\bibnamefont{Barkai}},
  \bibinfo{journal}{Phys. Rev. E} \textbf{\bibinfo{volume}{73}},
  \bibinfo{pages}{016125} (\bibinfo{year}{2006}).

\bibitem[{\citenamefont{van Kampen}(2001)}]{vanKampen}
\bibinfo{author}{\bibfnamefont{N.}~\bibnamefont{van Kampen}},
  \emph{\bibinfo{title}{Stochastic Processes in Physics and Chemistry}}
  (\bibinfo{publisher}{Elsevier}, \bibinfo{year}{2001}), \bibinfo{edition}{3rd}
  ed.

\bibitem[{\citenamefont{Munsky and Khammash}(2006)}]{Munsky:2005FSP}
\bibinfo{author}{\bibfnamefont{B.}~\bibnamefont{Munsky}} \bibnamefont{and}
  \bibinfo{author}{\bibfnamefont{M.}~\bibnamefont{Khammash}},
  \bibinfo{journal}{J. Chem. Phys.} \textbf{\bibinfo{volume}{124}},
  \bibinfo{pages}{044104} (\bibinfo{year}{2006}).

\bibitem[{\citenamefont{Burrage et~al.}(2006)\citenamefont{Burrage, Hegland,
  Macnamara, and Sidje}}]{Burrage:2006}
\bibinfo{author}{\bibfnamefont{K.}~\bibnamefont{Burrage}},
  \bibinfo{author}{\bibfnamefont{M.}~\bibnamefont{Hegland}},
  \bibinfo{author}{\bibfnamefont{S.}~\bibnamefont{Macnamara}},
  \bibnamefont{and} \bibinfo{author}{\bibfnamefont{R.}~\bibnamefont{Sidje}},
  \bibinfo{journal}{Proc. of The A.A.Markov 150th Anniversary Meeting} pp.
  \bibinfo{pages}{21--37} (\bibinfo{year}{2006}).

\bibitem[{\citenamefont{Munsky and Khammash}(2007)}]{Munsky:2007mtsFSP}
\bibinfo{author}{\bibfnamefont{B.}~\bibnamefont{Munsky}} \bibnamefont{and}
  \bibinfo{author}{\bibfnamefont{M.}~\bibnamefont{Khammash}},
  \bibinfo{journal}{J. Comp. Phys.} \textbf{\bibinfo{volume}{226}},
  \bibinfo{pages}{818} (\bibinfo{year}{2007}).

\bibitem[{\citenamefont{Peles et~al.}(2006)\citenamefont{Peles, Munsky, and
  Khammash}}]{Slaven:2006JCP}
\bibinfo{author}{\bibfnamefont{S.}~\bibnamefont{Peles}},
  \bibinfo{author}{\bibfnamefont{B.}~\bibnamefont{Munsky}}, \bibnamefont{and}
  \bibinfo{author}{\bibfnamefont{M.}~\bibnamefont{Khammash}},
  \bibinfo{journal}{J. Chem. Phys.} \textbf{\bibinfo{volume}{125}},
  \bibinfo{pages}{204104} (\bibinfo{year}{2006}).

\bibitem[{\citenamefont{Munsky and
  Khammash}(2008{\natexlab{a}})}]{Munsky:2008IEEE}
\bibinfo{author}{\bibfnamefont{B.}~\bibnamefont{Munsky}} \bibnamefont{and}
  \bibinfo{author}{\bibfnamefont{M.}~\bibnamefont{Khammash}},
  \bibinfo{journal}{IEEE Trans. Automat. Contr./IEEE Trans. Circuits and
  Systems: Part 1} \textbf{\bibinfo{volume}{52}}, \bibinfo{pages}{201}
  (\bibinfo{year}{2008}{\natexlab{a}}).

\bibitem[{\citenamefont{Gillespie}(1976)}]{Gillespie:1976}
\bibinfo{author}{\bibfnamefont{D.~T.} \bibnamefont{Gillespie}},
  \bibinfo{journal}{J. Comp. Phys.} \textbf{\bibinfo{volume}{22}},
  \bibinfo{pages}{403} (\bibinfo{year}{1976}).

\bibitem[{\citenamefont{Gillespie}(2001)}]{Gillespie:2001}
\bibinfo{author}{\bibfnamefont{D.~T.} \bibnamefont{Gillespie}},
  \bibinfo{journal}{J. Chem. Phys.} \textbf{\bibinfo{volume}{115}},
  \bibinfo{pages}{1716} (\bibinfo{year}{2001}).

\bibitem[{\citenamefont{Cao et~al.}(2005)\citenamefont{Cao, Gillespie, and
  Petzold}}]{Petzold:2005}
\bibinfo{author}{\bibfnamefont{Y.}~\bibnamefont{Cao}},
  \bibinfo{author}{\bibfnamefont{D.}~\bibnamefont{Gillespie}},
  \bibnamefont{and} \bibinfo{author}{\bibfnamefont{L.}~\bibnamefont{Petzold}},
  \bibinfo{journal}{J. Chem. Phys.} \textbf{\bibinfo{volume}{122}},
  \bibinfo{pages}{014116} (\bibinfo{year}{2005}).

\bibitem[{\citenamefont{Munsky and
  Khammash}(2008{\natexlab{b}})}]{Munsky:2008IET}
\bibinfo{author}{\bibfnamefont{B.}~\bibnamefont{Munsky}} \bibnamefont{and}
  \bibinfo{author}{\bibfnamefont{M.}~\bibnamefont{Khammash}},
  \bibinfo{journal}{IET Systems Biology} \textbf{\bibinfo{volume}{2}},
  \bibinfo{pages}{323} (\bibinfo{year}{2008}{\natexlab{b}}).

\bibitem[{\citenamefont{Dellago et~al.}(1998)\citenamefont{Dellago, Bolhuis,
  Csajka, and Chandler}}]{Dellago:1998}
\bibinfo{author}{\bibfnamefont{C.}~\bibnamefont{Dellago}},
  \bibinfo{author}{\bibfnamefont{P.}~\bibnamefont{Bolhuis}},
  \bibinfo{author}{\bibfnamefont{F.}~\bibnamefont{Csajka}}, \bibnamefont{and}
  \bibinfo{author}{\bibfnamefont{D.}~\bibnamefont{Chandler}},
  \bibinfo{journal}{J. Chem. Phys} \textbf{\bibinfo{volume}{108}},
  \bibinfo{pages}{1964} (\bibinfo{year}{1998}).

\bibitem[{\citenamefont{Faradjian and Elber}(2004)}]{Faradjian:2004}
\bibinfo{author}{\bibfnamefont{A.}~\bibnamefont{Faradjian}} \bibnamefont{and}
  \bibinfo{author}{\bibfnamefont{R.}~\bibnamefont{Elber}}, \bibinfo{journal}{J.
  Chem. Phys.} \textbf{\bibinfo{volume}{120}}, \bibinfo{pages}{10880}
  (\bibinfo{year}{2004}).

\bibitem[{\citenamefont{Moroni et~al.}(2004)\citenamefont{Moroni, Bolhuis, and
  van Erp.}}]{Moroni:2004}
\bibinfo{author}{\bibfnamefont{D.}~\bibnamefont{Moroni}},
  \bibinfo{author}{\bibfnamefont{P.}~\bibnamefont{Bolhuis}}, \bibnamefont{and}
  \bibinfo{author}{\bibfnamefont{T.}~\bibnamefont{van Erp.}},
  \bibinfo{journal}{J. Chem. Phys.} \textbf{\bibinfo{volume}{120}},
  \bibinfo{pages}{4055} (\bibinfo{year}{2004}).

\bibitem[{\citenamefont{van Erp. and Bolhuis}(2005)}]{Erp:2005}
\bibinfo{author}{\bibfnamefont{T.}~\bibnamefont{van Erp.}} \bibnamefont{and}
  \bibinfo{author}{\bibfnamefont{P.}~\bibnamefont{Bolhuis}},
  \bibinfo{journal}{J. Comp. Phys.} \textbf{\bibinfo{volume}{205}},
  \bibinfo{pages}{157} (\bibinfo{year}{2005}).

\bibitem[{\citenamefont{Allen and Frenkel}(2006)}]{Allen:2006JCPA}
\bibinfo{author}{\bibfnamefont{R.}~\bibnamefont{Allen}} \bibnamefont{and}
  \bibinfo{author}{\bibfnamefont{P.}~\bibnamefont{Frenkel},
  \bibfnamefont{D.~Rein ten~Wolde}}, \bibinfo{journal}{J. Chem. Phys.}
  \textbf{\bibinfo{volume}{124}}, \bibinfo{pages}{024102}
  (\bibinfo{year}{2006}).

\bibitem[{\citenamefont{Bel et~al.}(2009)\citenamefont{Bel, Munsky, and
  Nemenman}}]{BMN:2009}
\bibinfo{author}{\bibfnamefont{G.}~\bibnamefont{Bel}},
  \bibinfo{author}{\bibfnamefont{B.}~\bibnamefont{Munsky}}, \bibnamefont{and}
  \bibinfo{author}{\bibfnamefont{I.}~\bibnamefont{Nemenman}},
  \bibinfo{journal}{Submitted}  (\bibinfo{year}{2009}).

\end{thebibliography}

\end{document}